# Surface lattice solitons in diffusive nonlinear media

Yaroslav V. Kartashov[1], Victor A. Vysloukh[2], and Lluis Torner[1]

[1]*ICFO-Institut de Ciencies Fotoniques, and Universitat Politecnica de Catalunya, Mediterranean Technology Park, 08860 Castelldefels (Barcelona), Spain*

[2]*Departamento de Física y Matemáticas, Universidad de las Americas – Puebla, Puebla 72820, Mexico*

We address the properties of surface solitons supported by optical lattices imprinted in photorefractive media with asymmetric diffusion nonlinearity. Such solitons exist only in finite gaps of lattice spectrum. In contrast to latticeless geometries, where surface waves exist only when nonlinearity deflects light towards the material surface, the surface lattice solitons exist in settings where diffusion would cause beam bending against the surface.

*OCIS codes: 190.0190, 190.6135*

Several optical materials exhibit intrinsically asymmetric nonlinear responses, that cause the development of strong shape asymmetries or even self-bending of light beams. Among such materials are photorefractive crystals [1-6] with a diffusion nonlinearity. The diffusion nonlinearity can cause substantial transverse beam displacements when acting in combination with drift nonlinearity in biased crystals. In the presence of interfaces diffusion nonlinearity results in the formation of localized surface states when bending toward the interface is compensated by total internal reflection [7-10]. Such surface waves are an example of nonlinear surface waves observed at the interface of natural uniform materials [11]. Combining diffusion and drift nonlinearity may also result in surface wave formation [12].

Surface waves can also be excited at moderate powers at the interface between uniform medium and semi-infinite lattice [13,14]. They were found not only in focusing, but also in defocusing media [15-18], and in both one- and two-dimensional [19,20] geometries. To date, surface lattice solitons were studied mostly in materials featuring symmetric local nonlinear response. The impact of the asymmetric nonlocal nonlinear response on lattice solitons was studied only in infinite periodic structures [21,22]. In this Letter we address finite lattices, and reveal that nonlocal asymmetric diffusion nonlinearity can result in surface soliton formation in finite gaps in lattice spectrum. In contrast to the case of interface of uniform diffusive medium, such solitons exist when diffusion nonlinearity would cause light bending not only toward the lattice edge but also when it bends light against it.

The propagation of light beam along the $\xi$ axis in the vicinity of the interface of a semi-infinite optical lattice imprinted in an unbiased photorefractive crystal with intrinsic diffusion nonlinearity can be described by the nonlinear Schrödinger equation

$$i\frac{\partial q}{\partial \xi} = -\frac{1}{2}\frac{\partial^2 q}{\partial \eta^2} + \frac{\mu q}{1+S|q|^2}\frac{\partial |q|^2}{\partial \eta} - pR(\eta)q, \qquad (1)$$

where the transverse $\eta$ and longitudinal $\xi$ coordinates are scaled to the beam width $x_0$ and diffraction length $k_0 x_0^2$, respectively; $q = k_0 n_0 (x_0 k_b T r_e / 2eI_d)^{1/2} A$ is the complex field amplitude; $k_b$ is the Boltzmann constant; $T$ is the temperature; $r_e$ is the electro-optic coefficient; $I_d$ is the dark irradiance; $e$ is the charge of free carriers; $n_0$ is the refractive index; $S = 2e/k_0^2 x_0 n_0^2 k_b T r_e$ is the saturation parameter; the parameter $p = k_0^2 x_0^2 \delta n / n_0$ is proportional to the depth of the refractive index modulation $\delta n$; the lattice profile is described by the function $R(\eta) = 0$ for $\eta < 0$ and $R(\eta) = [1-\cos(\Omega \eta)]/2$ for $\eta \geq 0$. Such refractive index landscapes can be fabricated in suitable materials [14,17,18]. When $\mu = 1$, a light beam launched inside the lattice tends to self-bend towards the bulk lattice, while when $\mu = -1$ light tends to self-bend toward the lattice edge. Here we set $\Omega = 2$ and $S = 0.5$. The Eq. (1) conserves the energy flow $U = \int_{-\infty}^{\infty} |q|^2 \, d\eta$.

We search for lattice surface soliton solutions in the form $q(\eta,\xi) = w(\eta)\exp(ib\xi)$. Such states exist only for $b$ values falling into gaps of the lattice spectrum [Fig. 1(a)]. The tails of surface solitons decay in the uniform medium only when $b \geq 0$. These constraints determine approximately the domain of existence of surface solitons in diffusive medium. Surprisingly, such solitons can exist only in finite gaps of the lattice spectrum. Representative profiles of surface solitons belonging to the first gap are shown in Fig. 2.

In contrast to the case of uniform media, in the lattice surface solitons exist for both signs of $\mu$. The intensity maxima in the first lattice channel are shifted in opposite directions for opposite sings of $\mu$ [Fig. 2(a)], moreover, when $\mu = 1$ the soliton localization inside the lattice is lower than when $\mu = -1$. This is because in the former case Bragg reflection from the periodic structure responsible for near-surface localization counteracts also the self-bending against the surface.

There exist both lower and upper cutoffs on $b$. When $\mu = -1$ the lower cutoff for solitons from the first gap coincides with the lower gap edge at $p \geq p_{\text{cr}}^1$, while at $p_{\text{cr}}^1 > p \geq p_{\text{cr}}^2$ one has $b_{\text{low}} = 0$ [Fig. 1(c)]. In this case $p_{\text{cr}}^1 \approx 1.78$ and $p_{\text{cr}}^2 \approx 0.66$ correspond to the lattice depths at which the lower and upper gap edges cross the line $b = 0$. The upper cutoff $b_{\text{upp}}$ is a bit smaller, but still very close to upper gap edge. For $p > p_{\text{cr}}^1$ at $b \to b_{\text{low}}$ and at $b \to b_{\text{upp}}$ surface solitons largely expand into the lattice region, but are well localized in the uniform medium [Figs. 2(b) and 2(c)]. In contrast, at $p_{\text{cr}}^1 > p \geq p_{\text{cr}}^2$ for $b \to b_{\text{low}}$ surface solitons feature almost linearly decreasing tail inside uniform medium, but remain localized in the lattice [Fig. 2(d)]. The energy flow is a nonmonotonic function of $b$: It decreases with $b$ in most part of the existence domain, but at $b \to b_{\text{low}}$ the derivative $dU/db$ becomes positive, so that surface solitons exist only above a certain energy flow threshold [Fig. 2(b)]. The existence of such a threshold is entirely a surface effect.

We found that at $\mu = 1$ the domain of existence of surface solitons is much narrower than when $\mu = -1$ [Fig. 1(c)]. This is because lattice can not compensate the effects of self-bending for beams with sufficiently high amplitudes, and deeper lattices are necessary to support surface solitons with higher peak intensities. As a result, the domain of existence for surface solitons shrinks with decreasing lattice depth and such solitons cease to exist for $p < p_{\text{cr}}^3 \approx 2.61$, which exceeds $p_{\text{cr}}^1, p_{\text{cr}}^2$. The domain of existence for surface solitons for $\mu = 1$ is located closer to the upper gap edge [Fig. 1(c)]. When $b \to b_{\text{upp}}$ surface solitons strongly penetrate into the lattice [Fig. 2(e)], while at $b \to b_{\text{low}}$ an increase of the energy flow is accompanied by a gradual equilibration of amplitudes of oscillations in the first and second lattice channels [Fig. 2(f)]. In this case the tangential line to $U(b)$ curve becomes vertical in both cutoffs [Fig. 1(b)]. Surface lattice solitons for $\mu = 1$ also exist only above a threshold energy flow. We also found surface solitons with similar properties in other finite gaps of lattice spectrum for both signs of $\mu$.

A linear stability analysis for perturbed solutions $q = (w + u + iv)\exp(ib\xi)$ (here $u, v$ are real and imaginary parts of perturbation that can grow with complex rate $\delta = \delta_r + i\delta_i$ upon propagation) of Eq. (1) has shown that surface solitons in diffusive medium are stable in a largest part of their existence domain for both signs of $\mu$. Only close to the upper cutoff one encounters a domain of exponential instability, which coincides with the region where $dU/db \geq 0$ [see Fig. 1(d) for dependence $\delta_r(b)$ for $\mu=1$]. Such instability domain exists for both signs of $\mu$. For $b$ values close to the lower gap edge surface solitons suffer from very weak oscillatory instabilities typical for gap solitons. Figures 3(a) and 3(b) show stable propagation of perturbed surface solitons at $\mu=\pm 1$.

Importantly, the excitation dynamics for surface solitons out of arbitrary inputs strongly differs for opposite signs of $\mu$, since for $\mu=-1$ the bending towards the lattice edge facilitates surface wave formation, while for $\mu=1$ light tends to drift into the lattice bulk. Thus, higher energy flows are required for excitation of surface solitons when $\mu=1$. This is illustrated in Figs. 3(c), 3(d) where a Gaussian beam $q|_{\xi=0} = \exp[-(\eta-\pi/\Omega)^2]$, whose energy flow far exceeds the thresholds for surface wave formation for both signs of $\mu$, excites surface wave for $\mu=-1$, but diffracts when $\mu=1$. In the latter case the input energy flow has to be further increased to achieve surface soliton formation. Surface solitons may be excited with input beams carrying much higher energies than the energy flows of surface solitons in the vicinity of lower cutoff. In this case, despite a strong radiation, a share of input energy flow usually remains trapped in the near-surface channel. For $p<p_{cr}^2$ when $\mu=-1$ and for $p<p_{cr}^3$ when $\mu=1$, surface soliton formation is not possible, irrespectively of the shape and energy flow of input beam.

Summarizing, we introduced gap surface solitons supported by asymmetric nonlocal diffusion nonlinearity at the interface of optical lattices. Such solitons exist for both sings of the diffusive nonlinearity, in contrast to previous findings reported for latticeless geometries.

# References with titles

# References without titles

# Figure captions

Figure 1. (a) Floquet-Bloch spectrum of infinite lattice. Gray regions show bands, white regions correspond to gaps. (b) Energy flow versus propagation constant for $\mu=1$ (red line) and $\mu=-1$ (black line) at $p=6$. (c) Domain of existence of surface solitons on the $(p,b)$ plane for $\mu=1$ (red line) and $\mu=-1$ (black line). (d) Real part of perturbation growth rate versus propagation constant at $p=6$, $\mu=1$. In all cases $\Omega=2$.

Figure 2. Profiles of surface solitons at (a) $b=1.9$, $p=3.5$, $\mu=-1$ (black line) and $\mu=1$ (red line), (b) $b=0.62$, $p=3.5$, $\mu=-1$, (c) $b=2.27$, $p=3.5$, $\mu=-1$, (d) $b=0.1$, $p=1.5$, $\mu=-1$, (e) $b=3.53$, $p=5$, $\mu=1$, (f) $b=2.4$, $p=5$, $\mu=1$. In gray regions $R(\eta) \geq 1/2$, while in white regions $R(\eta) < 1/2$. In all cases $\Omega=2$.

Figure 3. Stable propagation of perturbed surface solitons corresponding to $b=1.9$, $p=3.5$, $\mu=-1$ (a) and $b=2.23$, $p=3.5$, $\mu=1$ (b). The broadband white noise was added into input field distributions. Dynamics of propagation of Gaussian beam launched into first channel of lattice with $p=3$ at $\mu=1$ (c) and $\mu=-1$ (d). In all cases field modulus distributions are shown, white dashed lines indicate interface position, and lattice frequency $\Omega=2$.

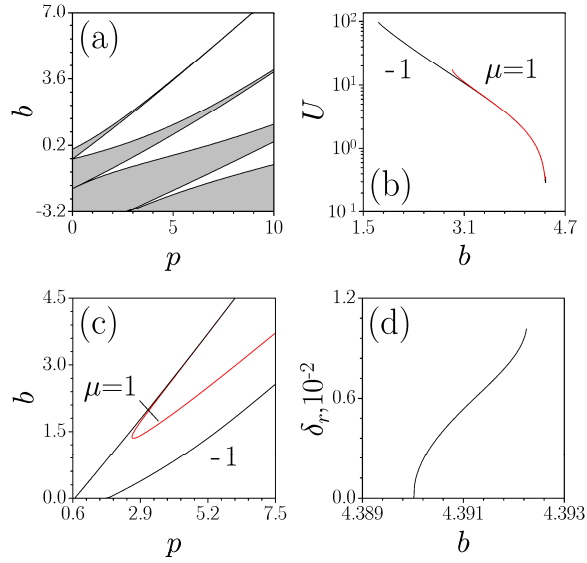

Figure 1. (a) Floquet-Bloch spectrum of infinite lattice. Gray regions show bands, white regions correspond to gaps. (b) Energy flow versus propagation constant for $\mu=1$ (red line) and $\mu=-1$ (black line) at $p=6$. (c) Domain of existence of surface solitons on the $(p,b)$ plane for $\mu=1$ (red line) and $\mu=-1$ (black line). (d) Real part of perturbation growth rate versus propagation constant at $p=6$, $\mu=1$. In all cases $\Omega=2$.

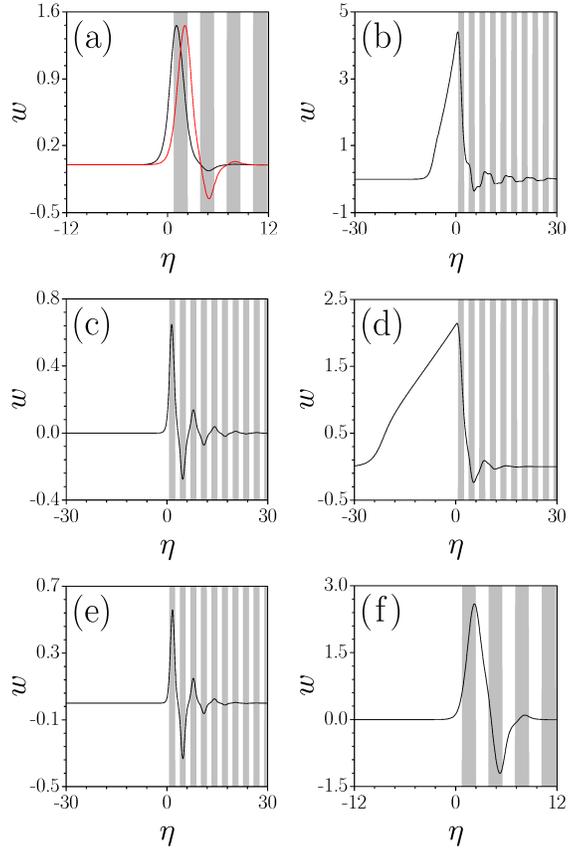

Figure 2. Profiles of surface solitons at (a) $b = 1.9$, $p = 3.5$, $\mu = -1$ (black line) and $\mu = 1$ (red line), (b) $b = 0.62$, $p = 3.5$, $\mu = -1$, (c) $b = 2.27$, $p = 3.5$, $\mu = -1$, (d) $b = 0.1$, $p = 1.5$, $\mu = -1$, (e) $b = 3.53$, $p = 5$, $\mu = 1$, (f) $b = 2.4$, $p = 5$, $\mu = 1$. In gray regions $R(\eta) \geq 1/2$, while in white regions $R(\eta) < 1/2$. In all cases $\Omega = 2$.

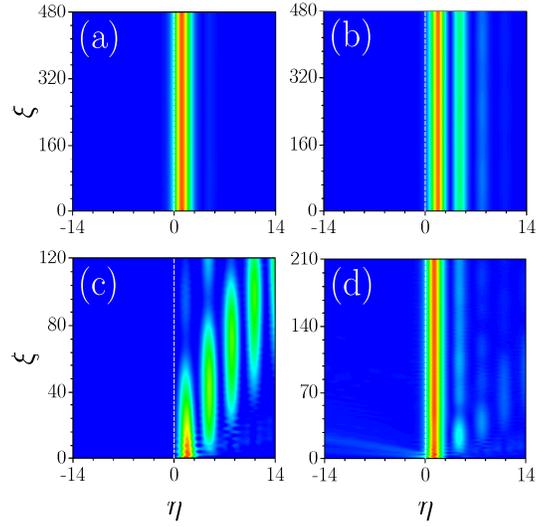

Figure 3. Stable propagation of perturbed surface solitons corresponding to $b = 1.9$, $p = 3.5$, $\mu = -1$ (a) and $b = 2.23$, $p = 3.5$, $\mu = 1$ (b). The broadband white noise was added into input field distributions. Dynamics of propagation of Gaussian beam launched into first channel of lattice with $p = 3$ at $\mu = 1$ (c) and $\mu = -1$ (d). In all cases field modulus distributions are shown, white dashed lines indicate interface position, and lattice frequency $\Omega = 2$.